\documentclass[conference]{IEEEtran}
\IEEEoverridecommandlockouts
\usepackage{cite}
\usepackage{amsmath,amssymb,amsfonts}
\usepackage{algorithmic}
\usepackage{graphicx}
\usepackage{textcomp}
\usepackage{xcolor}
\def\BibTeX{{\rm B\kern-.05em{\sc i\kern-.025em b}\kern-.08em
    T\kern-.1667em\lower.7ex\hbox{E}\kern-.125emX}}

\usepackage{amsmath}
\usepackage{amsfonts}
\usepackage{amssymb}
\usepackage{graphicx}
\usepackage{graphics}
\usepackage{graphicx}
\usepackage{graphics}
\usepackage{graphics,color}
\usepackage{graphics,color}
\usepackage{url}
\usepackage{epstopdf}
\usepackage{tabularx}
\usepackage{tikz}
\usetikzlibrary{matrix}
\usepackage{cite}
\usepackage{ifthen}
\usepackage{times}
\usepackage{tabularx}
\usepackage{epstopdf}
\usepackage{bm}

\newcommand{\hide}[1]{\ifthenelse{\boolean{false}}{#1}{}}

\newtheorem{theorem}{{\bf Theorem}}
\newtheorem{lemma}{{\bf Lemma}}
\newtheorem{claim}{{\bf Claim}}

\newtheorem{corollary}{{\bf Corollary}}

\newtheorem{remark}{{\bf Remark}}

\newenvironment{proof}[1][Proof]{\begin{trivlist}
		\item[\hskip \labelsep {\bfseries #1}]}{\end{trivlist}}

\newtheorem{defn}{Definition}

\newcommand{\barr}{\begin{array}}
	\newcommand{\earr}{\end{array}}

\newcommand{\benum}{\begin{enumerate}}
	\newcommand{\eenum}{\end{enumerate}}

\newcommand{\bit}{\begin{itemize}}
	\newcommand{\eit}{\end{itemize}}

\newcommand{\bdes}{\begin{description}}
	\newcommand{\edes}{\end{description}}

\newcommand{\bfig}{\begin{figure}}
	\newcommand{\efig}{\end{figure}}

\newcommand{\bemq}{\begin{quote} \begin{em}}
		\newcommand{\eemq}{\end{em} \end{quote}}

\newcommand{\qed}{\hfill\ensuremath{\square}}%

\newcommand{\given}{\arrowvert}

\newcommand{\bt}{\begin{theorem}}
	\newcommand{\bl}{\begin{lemma}}
		\newcommand{\bc}{\begin{claim}}
			\newcommand{\bp}{\begin{Proposition}}
				\newcommand{\bcoro}{\begin{corollary}}
					\newcommand{\bres}{\begin{Result}}
						\newcommand{\brem}{\begin{Remark}}
							\newcommand{\et}{\end{theorem}}
						\newcommand{\el}{\end{lemma}}
					\newcommand{\ec}{\end{claim}}
				\newcommand{\ep}{\end{Proposition}}
			\newcommand{\ecoro}{\end{corollary}}
		\newcommand{\eres}{\end{Result}}
	\newcommand{\erem}{\end{Remark}}

\newcommand{\beq}{\begin{equation}}
	\newcommand{\eeq}{\end{equation}}

\newcommand{\UN}[1]{{\mathcal{V}}^{(N)}_{k}}

\newcommand{\mb}[1]{\mathbb{#1}}
\newcommand{\mf}[1]{\mathbf{#1}}
\newcommand{\mc}[1]{\mathcal{#1}}

\begin{document}
	
	\title{Quickest Change Detection in the Presence of  Transient Adversarial Attacks\\
		\thanks{This work was supported in part by the Army Research Laboratory under Cooperative Agreement W911NF-17-2-0196 (IoBT CRA).}
	}
	
	\author{\IEEEauthorblockN{Thirupathaiah Vasantam}
		\IEEEauthorblockA{\textit{Dept. of CICS} \\
			\textit{University of Massachusetts}\\
			Amherst, USA \\
			tvasantam@umass.edu}
		\and
		\IEEEauthorblockN{ Don Towsley}
		\IEEEauthorblockA{\textit{Dept. of CICS} \\
			\textit{University of Massachusetts}\\
			Amherst, USA \\
			towsley@cs.umass.edu}
		\and
		\IEEEauthorblockN{Venugopal V. Veeravalli}
\IEEEauthorblockA{\textit{Dept. of ECE} \\
			\textit{University of Illinois}\\
			Urbana-Champaign, USA \\
			vvv@illinois.edu}
			}
\maketitle


\begin{abstract}  
We study a monitoring system in which the distributions of sensors' observations change from a nominal distribution to an abnormal distribution in response to an adversary's presence. The system uses the quickest change detection procedure, the Shewhart rule, to detect the adversary that uses its resources to affect the abnormal distribution, so as to hide its presence. The metric of interest is the probability of missed detection within a predefined number of time-slots after the changepoint. Assuming that the adversary's resource constraints are known to the detector, we find the number of required sensors to make the worst-case probability of missed detection less than an acceptable level. The distributions of observations are assumed to be Gaussian, and the presence of the adversary affects their mean. We also provide simulation results to support our analysis.

\end{abstract}
\section{Introduction}
The problem of quick detection of changes in statistical properties of a random process appears in many applications, ranging from fault detection, radar signal processing, industrial quality control, and navigational systems \cite{Poor_book,Egea-Roca_2018, Blaise_2012,Blakhache_2000}. The existing works on this detection problem have focused on two types of models. The first is the quick change detection (QCD) problem, which deals with the case where the distribution changes from a nominal distribution to another distribution. The second is the transient change detection (TCD) problem, where the distribution changes from the nominal distribution to an abnormal distribution for a finite duration, and then it reverts back to the original nominal distribution. The problem of interest in this paper is a hybrid of the QCD and TCD problems. As in the QCD problem, the distribution of the observations changes persistently to a different distribution in response to an adversary; however, for a certain number of time-slots after the changepoint, the adversary controls the abnormal distribution by investing resources with the goal of hindering the change detection. 
 Once the adversary's resources expire, the distribution of the observations returns to the persistent abnormal distribution.

%
%
%

The main objective in the QCD problem is to minimize average detection delay, subject to certain false-alarm constraints.
In the Bayesian framework, under the assumption that the changepoint is a random variable with a known distribution (typically geometric), the optimization problem is to minimize the average detection delay with an upper bound constraint  on the probability of false-alarm\cite{Tartakovsky_bayesian}. When the changepoint is deterministic but unknown, the detection problem is studied using a minimax framework, where the objective is to minimize the worst-case average detection delay, subject to a lower bound constraint on the mean time to false-alarm\cite{Lorden,Pollak1985}. The Cumulative Sum (CUSUM) test introduced by Page~\cite{page}, was shown to achieve optimal performance in \cite{Moustakides1986}.

For many applications such as 
monitoring in global navigation satellite systems (GNSSs)\cite{Toledo_GNSS1}, monitoring in water distribution systems\cite{Blaise_2012}, monitoring in industrial processes\cite{Industrial_book}, it is of interest to detect abnormalities in a monitoring process with a certain tolerable detection delay, say $K$ time-slots after the changepoint, and if the detector announces a change in the distribution after $K$ time-slots then it is considered to be a missed detection. Hence the goal here is to minimize the probability of missed detection (PMD) in $K$ time-slots after the changepoint. Such a metric is commonly used in TCD problems, with $K$ corresponding to the duration of the transient.

In \cite{Moustakides_Shewhart}, it was shown that the Shewhart test\cite{shewhart} minimizes the PMD when $K=1$. The performance of CUSUM based algorithms for finite observation intervals was studied in \cite{Wang2005AVT,Wang_2000}. The performance of a window length-based CUSUM test was investigated in \cite{Fillatre_2015,Blaise_2012}, where they obtained some upper bounds on the PMD and proposed a numerical method to compute the PMD. Recently, some performance bounds for the finite moving average method were obtained in \cite{Egea-Roca_2018}.

We assume that the detector uses $M$ active sensors for detection, and that the adversary uses its resources to attack all of the active sensors. 
We find the least number of sensors that the detector must use to make the worst-case PMD in $K$ time-slots after the changepoint less than some acceptable level.

The rest of the paper is organized as follows. In Section~\ref{sec:model}, we introduce the model. In Section~\ref{sec:rule}, we define the stopping rule and provide some preliminary results. In Section~\ref{sec:results}, we give the main results on the worst-case PMD achievable by the adversary. In Section~\ref{sec:numerics}, we provide some numerical results to validate our analysis, and in Section~\ref{sec:conclusions}, we give some concluding remarks.


\section{Problem Model}
\label{sec:model}
Consider a system with $M$ sensors that take observations to detect an adversary's presence in the system.
Let $\mf{X}_n=(X_n^{(1)},\cdots,X_n^{(M)})$ where $X_n^{(i)}$ is the observation of sensor $i$ at time-slot $n$.
The elements of the sequence $\{\mf{X}_n\}_{n\geq 1}$ are assumed to be independent random variables. Furthermore, $\{X_n^{(i)}\}_{1\leq i\leq M}$ are assumed to be independent and identically distributed random variables for each $n$. We assume that the observations are real-valued and continuous, and have probability density functions (pdfs).
The pdf of $X_n^{(i)}$ is denoted by $p_{X,n}$. 
The nominal pdf of the system is denoted by $f_{0}$. In previous works, the distribution is assumed to change from a distribution with pdf $f_{0}$ to another distribution with pdf $f_{1}$ at a deterministic but unknown time instant referred to as the changepoint, denoted by $\nu$. In this paper,
we assume that an adversary changes the pdf $p_{X,n}$ from $f_{0}$ to $\psi_{\gamma(\theta)}$, where the parameter $\theta$ is chosen by the adversary and the function $\gamma(\theta)$ is defined as
\beq
\gamma(\theta)=\frac{\theta}{M}.
\eeq
 Here, the
parameter $\theta$ indicates the amount of resources invested by the adversary in each time-slot to hide its presence and we assume that it uses $\gamma(\theta)=\frac{\theta}{M}$ resources on each sensor. We denote the Kullback-Leibler (KL) divergence between the distributions $P$ and $Q$ with pdfs $p$ and $q$ by $D(p,q)$, where
\beq
D(p,q)=\int p(x)\log\left(\frac{p(x)}{q(x)}\right)\,dx.
\eeq
We make an
assumption that the KL divergence $D(\psi_{\gamma(\theta)},f_{0})$
decreases monotonically in $\theta$ and $\lim_{\theta \to\infty}D(\psi_{\gamma(\theta)},f_{0})=0$, indicating that $\psi_{\gamma(\theta)}$ is nearly the same as $f_{0}$ when $\theta$ is large.
If the adversary does not put any effort to hide its presence, then $\theta=0$. Also, we assume that $\psi_{\gamma(\infty)}=f_0$ and $\psi_{\gamma(0)}=f_1$.
Furthermore, the adversary is assumed to have a finite amount of resources to change the post-change distribution's pdf from $f_{1}$ to $\psi_{\gamma(\theta)}$, $\theta>0$, for a duration of $L_{\theta}$ time-slots after the changepoint.
Precisely, we consider the following framework: at time-slot $n$ the pdf $p_{X,n}$ is equal to $\psi_{\gamma(\theta)}$ where the function $\gamma(\theta)$ satisfies
\begin{equation}
	\label{eq:main}
	\gamma(\theta)=
	\begin{cases}
		\infty,&  \text{ for }1\leqslant n \leq \nu \\
		\frac{\theta}{M},& \text{ for } \nu+1\leq n \leq \nu+L_{\theta} \\
		0,& \text{ for } n \geq \nu+L_{\theta}+1,
	\end{cases}
\end{equation}
satisfying $\theta\in\mb{R}_+$ and $L_{\theta}\in\mb{Z}_+$ is a decreasing function of $\theta$. Let $A$ be the total amount of resources of the adversary and $\theta$ the amount of invested resources in each time-slot, then $L_{\theta}$ is the largest integer satisfying $L_{\theta}\leq \frac{A}{\theta}$.

Let $Q_{\theta}(K,M,A)$ be the PMD in $K$ time-slots after the changepoint, where $K$ is a fixed parameter with $K\geq L_{\theta}$.
An adversary's objective is to maximize its impact on the system 
by maximizing $Q_{\theta}(K,M,A)$. Hence, the adversary's objective is to find $\theta$ that maximizes $Q_{\theta}(K,M,A)$. 
We assume that the system operator uses a detector with the Shewhart stopping rule to detect changes in the distributions of the observations under the assumption that $f_{0}$ and $f_{1}$ are pre-change and post-change pdfs, respectively. The detector is unaware of the adversary's actions during the transient regime. However, under the assumption that the detector has information about the adversary's resources, we first compute $\theta$ that results in the worst-case PMD. We then find the number of sensors to be used by the detector to make the worst-case PMD less than a predefined parameter $\delta$.


In this paper, we study the scenario where the distributions are Gaussian and focus on the mean-change problem. However, the analysis can be generalized to other settings. For a random variable $Z_n$, we write the conditional probabilities and expectations as $\mb{P}_{\nu,\theta}(Z_n)$ and $\mb{E}_{\nu,\theta}[Z_n]$ when the changepoint is $\nu$ and each sensor's observation has pdf $\psi_{\gamma{(\theta)}}$ in time-slot $n$. Also, $\mb{P}_{\infty}(Z_n)$ and $\mb{E}_{\infty}[ Z_n ]$ are the conditional probabilities and expectations of $Z_n$ when the observations of sensors have nominal distributions in time-slot $n$.

\section{Stopping Rule}
\label{sec:rule}
%
%
%
%
%
%
%
%
In this section, we first define the stopping rule, and then we derive an expression for $Q_{\theta}(K,M,A)$.

Let $g_{\theta}$ be the joint pdf of $\mf{X}_n$ when the adversary uses $\theta$ resources in each time-slot in the transient regime after the changepoint and $g_{\infty}$ be the joint pre-change pdf of $\mf{X}_n$. Next, we define the stopping test. The detector is assumed to have information that the adversary has at most $A$ resources.
\begin{defn}{Shewhart test:}
  For given false-alarm probability $\alpha$ the stopping time $T$ is defined as
\beq
T\triangleq \inf\left\{n\geq 1: w(\mf{X}_n)\geq h' \right\},
\eeq
where $w(\mf{X}_n)=\sup_{\theta\in[0,A]}\log\left({\frac{g_{\theta}(\mf{X}_n)}{g_{\infty}(\mf{X}_n)}}\right)$ and $h'$ is selected such that
$
\mb{P}_{\infty}\left(w(\mf{X}_n)\geq h' \right)=\alpha.
$
\end{defn}

%

Now we use the assumption that the distributions of observations are Gaussian.
Let $\phi_{z}$ be the pdf of the Gaussian distribution $\mc{N}(\mu_z,1)$, where $z$ is a parameter that defines the mean $\mu_z$. The function $\mu_{z}$ decreases monotonically in $z$ satisfying $\lim_{z\to\infty}\mu_{z}=0$. Since the observations' distributions are Gaussian
\beq
\phi_{\gamma(\theta)}=\psi_{\gamma(\theta)},
\eeq
for $\theta \in\mb{R}_+$.
The nominal and the persistent abnormal pdfs are assumed be $f_{0}=\phi_{\infty}$ and $f_{1}=\phi_{0}$, respectively.

For the mean-change problem, it can be verified that the likelihood ratio $\log\left({\frac{g_{\theta}(\mf{X}_n)}{g_{\infty}(\mf{X}_n)}}\right)$ is a monotonically increasing function in $Y_n$ defined by
\beq
Y_n=\sum_{i=1}^M X^{(i)}_n.
\eeq

Hence the Shewhart stopping time $T$ simplifies to the following form:
\beq
\label{eq:shewhart}
T\triangleq \inf\left\{n\geq 1: Y_n\geq h \right\},
\eeq
where $h$ is selected such that
\beq
\mb{P}_{\infty}\left(Y_n\geq h \right)=\alpha.
\eeq


Let $q_{\theta}$ be the PMD in a time-slot given by
\beq
q_{\theta}=\mb{P}_{\nu,\theta}(Y_n<h).
\eeq
Then it can be shown that
\beq
q_{\theta}=\Phi_{\infty}(\Phi_{\infty}^{-1}(1-\alpha)-\sqrt{M}\mu_{\gamma(\theta)}),
\eeq
where, $\Phi_{\infty}(\cdot)$ is the cumulative distribution function (CDF) of $\mc{N}(0,1)$ and $\Phi_{\infty}^{-1}(\cdot)$ is the inverse function of $\Phi_{\infty}(\cdot)$.

We give an expression for $Q_{\theta}(K,M,A)$ in the following Lemma and skip the proof as it follows immediately.
\begin{lemma}
	For given $\alpha$ with $\mu_{\theta}>0$ and $\mu_0>0$, we obtain
	\beq
	\label{eq:mean_det}
	Q_{\theta}(K,M,A)=q_{\theta}^{L_{\theta}}\times q_{0}^{K-L_{\theta}}.
	\eeq
\end{lemma}
Note that the PMD $Q_{\theta}(K,M,A)$ decreases monotonically with the number of sensors $M$.

\begin{remark}
	Since $\mu_{\theta}$ decreases monotonically in $\theta$, we have that 
	\beq
	\label{eq:resource_effect}
	q_{\theta}>q_{0}.
	\eeq
	The above inequality reflects that the probability of missed detection in a single time-slot increases with the amount of invested resources of the adversary.
\end{remark}

Note that although $q_{\theta}$ increases with $\theta$, the length of the transient duration $L_{\theta}$ decreases with $\theta$. Hence, the adversary chooses $\theta$ so as to maximize $Q_{\theta}(K,M,A)$.

Next, we show that if the adversary decides to invest $\theta$ resources in a time-slot, then it must invest $\frac{\theta}{M}$ resources on each sensor to maximize the probability of missed
detection. Assuming that the detector uses the stopping test $T$ defined in \eqref{eq:shewhart}, suppose $q^{(1)}_{\theta}$ is the probability of missed detection per time-slot when the adversary spends $\theta_i$ resources on Sensor $i$ ($1\leq i\leq M$) such that $\sum_{i=1}^M\theta_i=\theta$. Similarly, suppose $q^{(2)}_{\theta}$ is the probability of missed detection per time-slot when the adversary spends $\frac{\theta}{M}$ resources on each sensor. 

\begin{lemma}
	We show that
	\beq
	q^{(2)}_{\theta}\geq q^{(1)}_{\theta}.
	\eeq
	\end{lemma}
\begin{proof}
	It can be shown that
	\beq
	q^{(1)}_{\theta}=\Phi_{\infty}\left(\Phi_{\infty}^{-1}(1-\alpha)-\frac{1}{\sqrt{M}}\sum_{i=1}^M\mu_{\theta_i}\right)
	\eeq
	and
	\beq
	q^{(2)}_{\theta}=\Phi_{\infty}\left(\Phi_{\infty}^{-1}(1-\alpha)-\sqrt{M}\mu_{\frac{\theta}{M}}\right).
	\eeq
	To prove the result, we need to show that
	\beq
	\sqrt{M}\mu_{\frac{\theta}{M}}\leq \frac{1}{\sqrt{M}}\sum_{i=1}^M\mu_{\theta_i}.
	\eeq
	That is, we need to show
	\beq
	\mu_{\frac{\theta}{M}}\leq \frac{1}{M}\sum_{i=1}^M\mu_{\theta_i}.
	\eeq
	Since $\mu_{\theta}$ is strictly convex in $\theta$, from Jensen's inequality\cite{Poor_book}, we obtain
	\beq
	\frac{1}{M}\sum_{i=1}^M\mu_{\theta_i}\geq \mu_{\sum_{i=1}^M\frac{\theta_i}{M}}=\mu_{\frac{\theta}{M}}.
	\eeq
	This completes the proof.\qed
	\end{proof}
From the above result, the detector can assume that the adversary spends resources equally on each sensor. 

\section{Main Results}
\label{sec:results}
In this section, we show that there exists a unique $\theta$ that maximizes $Q_{\theta}(K,M,A)$ under certain sufficient conditions
for given $K$ and $A$. In our analysis, we study various types of functions for $L_{\theta}$. Although $L_{\theta}$ is an integer value, we allow a real valued $L_{\theta}$ to simplify our analysis.

Next, we define some useful notation. Define $r_{\theta}$ as
\beq
r_{\theta}=\log(Q_{\theta}(K,M,A)).
\eeq
Then we have
\begin{align}
	r_{\theta}
	&=L_{\theta}\log(q_{\theta})+(K-L_{\theta})\log(q_{0})\nonumber\\
	&=L_{\theta}\log\left(\frac{q_{\theta}}{q_{0}}\right)+K\log(q_{0}).\label{eq:rtheta}
\end{align}

The optimal $\theta$ that maximizes $Q_{\theta}(K,M,A)$ is a solution of the following equation
\beq
\label{eq:rderiv2}
\frac{d r_{\theta}}{d \theta}=\frac{d L_{\theta}}{d \theta}\log\left(\frac{q_{\theta}}{q_{0}}\right)+\frac{L_{\theta}}{q_{\theta}}\frac{d q_{\theta}}{d \theta}=0.
\eeq

Next, we establish the following preliminary result.
\begin{lemma}
	\label{thm:gauss}
	Let $\phi_{\infty}(\cdot)$ and $\Phi_{\infty}(\cdot)$ be the pdf and CDF of a $\mc{N}(0,1)$ random variable. Then, for $x\in\mb{R}$
	\beq
	\frac{\phi_{\infty}(x)}{\Phi_{\infty}(x)}+x>0.
	\eeq
\end{lemma}
\begin{proof}
	 For $x\in\mb{R}_+$, clearly, we have $\frac{\phi_{\infty}(x)}{\Phi_{\infty}(x)}+x>0$. We next show that even for $x<0$, the result $\frac{\phi_{\infty}(x)}{\Phi_{\infty}(x)}+x>0$ holds. For $y>0$, if $X$ is an $\mc{N}(0,1)$ random variable, then we can write
	\begin{align}
		\mb{E}(X\given X>y)&=\frac{1}{1-\Phi_{\infty}(y)}\int_{z=y}^{\infty}z  \phi_{\infty}(z)\,dz\\
		&=\frac{\phi_{\infty}(y)}{1-\Phi_{\infty}(y)}.
	\end{align}
	Since $\phi_{\infty}(y)=\phi_{\infty}(-y)$ and $\Phi_{\infty}(y)+\Phi_{\infty}(-y)=1$, we get
	\beq
	\mb{E}(X\given X>y)
	=\frac{\phi_{\infty}(-y)}{\Phi_{\infty}(-y)}.
	\eeq
	Due to the fact that $\mb{E}(X\given X>y)>y$, we obtain
	\beq
	\frac{\phi_{\infty}(-y)}{\Phi_{\infty}(-y)}>y.
	\eeq
	Hence, $\frac{\phi_{\infty}(-y)}{\Phi_{\infty}(-y)}+(-y)>0$ for $y>0$. \qed
	\end{proof}

Next, we will show that under certain conditions, we obtain $\frac{d^2 r_{\theta}}{d^2 \theta}<0$, which implies the uniqueness of a critical point. We use the notation $x^*=\Phi_{\infty}^{-1}(1-\alpha)$.
\begin{theorem}
	\label{thm:concave_L}
	If $\mu_{\theta}$ is strictly convex and $L_{\theta}$ is strictly concave, then there exists a unique critical point for $r_{\theta}$ denoted by $\theta^*$. Furthermore, $r_{\theta}$ increases with $\theta$ up to $\theta^*$ and then it decreases with $\theta$.
	The critical point $\theta^*$ is found by solving \eqref{eq:rderiv2}.
\end{theorem}
\begin{proof}
From \eqref{eq:rderiv2}, it can be shown that
\begin{multline}
\frac{d^2 r_{\theta}}{d^2 \theta}=\frac{d^2 L_{\theta}}{d^2 \theta}\log\left(\frac{q_{\theta}}{q_{0}}\right)+2\frac{d L_{\theta}}{d \theta}   \frac{1}{q_{\theta}}\frac{d q_{\theta}}{d \theta}\\
 -\frac{L_{\theta}}{q_{\theta}^2}\left(\frac{d q_{\theta}}{d \theta}\right)^2
+\frac{L_{\theta}}{q_{\theta}}\frac{d^2 q_{\theta}}{d^2 \theta}.
\end{multline}
Also, we can show that
\beq
\label{eq:qderiv}
\frac{d q_{\theta}}{d \theta}=\frac{1}{\sqrt{M}}\phi_{\infty}(x^*-\sqrt{M}\mu_{\gamma(\theta)})\left(\frac{-d \mu_{\gamma(\theta)}}{d \theta}\right)
\eeq
and
\begin{multline}
	\label{eq:qderiv2}
\frac{d^2 q_{\theta}}{d^2 \theta}=(x^*-\sqrt{M}\mu_{\gamma(\theta)})\frac{1}{M}\phi_{\infty}(x^*-\sqrt{M}\mu_{\gamma(\theta)})\left(\frac{-d \mu_{\gamma(\theta)}}{d \theta}\right)^2\\
+\frac{1}{M\sqrt{M}}\phi_{\infty}(x^*-\sqrt{M}\mu_{\gamma(\theta)})\left(\frac{-d^2 \mu_{\gamma(\theta)}}{d^2 \theta}\right).
\end{multline}

By using \eqref{eq:qderiv}-\eqref{eq:qderiv2} and \eqref{eq:qderiv2}, we obtain
\begin{multline}
	\label{eq:rderivnew}
	\frac{d^2 r_{\theta}}{d^2 \theta}=\frac{d^2 L_{\theta}}{d^2 \theta}\log\left(\frac{q_{\theta}}{q_{0}}\right)\\
	+2\frac{d L_{\theta}}{d \theta}   \frac{1}{q_{\theta}}\frac{1}{\sqrt{M}}\phi_{\infty}(x^*-\sqrt{M}\mu_{\gamma(\theta)})\left(\frac{-d \mu_{\gamma(\theta)}}{d \theta}\right)\\
	-\frac{L_{\theta}}{q_{\theta}}\left(\frac{1}{\sqrt{M}}\phi_{\infty}(x^*-\sqrt{M}\mu_{\gamma(\theta)})\left(\frac{-d \mu_{\gamma(\theta)}}{d \theta}\right)\right)^2\\
	\times\left(\frac{1}{q_{\theta}}\phi_{\infty}(x^*-\sqrt{M}\mu_{\gamma(\theta)})+\frac{1}{\sqrt{M}}(x^*-\sqrt{M}\mu_{\gamma(\theta)})\right)\\
	+\frac{L_{\theta}}{q_{\theta}}\frac{1}{M\sqrt{M}}\phi_{\infty}(x^*-\sqrt{M}\mu_{\gamma(\theta)})\left(\frac{-d^2 \mu_{\gamma(\theta)}}{d^2 \theta}\right).
\end{multline}

Since $\frac{dL_{\theta}}{d\theta}<0$, $\frac{d\mu_{\theta}}{d\theta}<0$, $\frac{d^2L_{\theta}}{d^2\theta}<0$, and $\frac{d^2\mu_{\theta}}{d^2\theta}>0$, the first, second, and fourth terms on the right side of \eqref{eq:rderivnew} are negative. To complete the proof, it remains to show that the third term is also negative. For this, it suffices to show that $\frac{1}{q_{\theta}}\phi_{\infty}(x^*-\sqrt{M}\mu_{\gamma(\theta)})+\frac{1}{\sqrt{M}}(x^*-\sqrt{M}\mu_{\gamma(\theta)})>0$. This holds if
$(x^*-\sqrt{M}\mu_{\gamma(\theta)})>0$. On the other hand, if $(x^*-\sqrt{M}\mu_{\gamma(\theta)})<0$, then 
\beq
\frac{1}{\sqrt{M}}(x^*-\sqrt{M}\mu_{\gamma(\theta)})>(x^*-\sqrt{M}\mu_{\gamma(\theta)}).
\eeq
Finally, from Lemma~\ref{thm:gauss}, we get $\frac{1}{q_{\theta}}\phi_{\infty}(x^*-\sqrt{M}\mu_{\gamma(\theta)})+\frac{1}{\sqrt{M}}(x^*-\sqrt{M}\mu_{\gamma(\theta)})>0$.
\qed
\end{proof}

Note that if $L_{\theta}$ is strictly convex, then the first term on the right side of \eqref{eq:rderivnew} is positive. Hence, the results of Theorem~\ref{thm:concave_L} cannot be extended to the case if $L_{\theta}$ is convex. However, we will show that the uniqueness of a critical point holds for certain types of convex functions for $L_{\theta}$. We study the following two cases: $L_{\theta}=\frac{A}{\theta}$ and $L_{\theta}=ae^{-\theta}$. Here, $a$ is chosen such that the total invested resources is less than $A$.


\begin{theorem}
	For $L_{\theta}=\frac{A}{\theta}$, if $\mu_{\theta}$ is strictly convex, then there exists a unique critical point for $r_{\theta}$ denoted by $\theta^*$. Furthermore, $r_{\theta}$ increases with $\theta$ up to $\theta*$ and then it decreases with $\theta$.
	The critical point $\theta^*$ satisfies
	\begin{align}
		\label{eq:critical_mean}
		\theta^*&=-\frac{\sqrt{M}q_{\theta^*}\log(\frac{q_{\theta^*}}{q_{0}})}{\phi_{\infty}(x^*-\sqrt{M}\mu_{\gamma(\theta^*)})\frac{d \mu_{\gamma(\theta^*)}}{d \theta}}.
	\end{align}
\end{theorem}
\begin{proof}
	
	By using $L_{\theta}=\frac{A}{\theta}$ and \eqref{eq:rderiv2}, we obtain that a critical point satisfies
	\beq
	\frac{-A}{ \theta^2}\log\left(\frac{q_{\theta}}{q_{0}}\right)+\frac{A}{\theta q_{\theta}}\left(\frac{d q_{\theta}}{d \theta}\right)=0.
	\eeq
	After simplifications we obtain
	\beq
	-\log\left(\frac{q_{\theta}}{q_{0}}\right)+\frac{\theta}{q_{\theta}}\left(\frac{d q_{\theta}}{d \theta}\right)=0.
	\eeq
	Let $b_{\theta}$ be defined as
	\beq
	b_{\theta}=-\log\left(\frac{q_{\theta}}{q_{0}}\right)+\frac{\theta}{q_{\theta}}\left(\frac{d q_{\theta}}{d \theta}\right).
	\eeq
	Then every critical point satisfies $b_{\theta}=0$. We will show that $\frac{d\,b_{\theta}}{d\,\theta}<0$ which implies that there exists a unique solution to the equation $b_{\theta}=0$. It can be verified that
	\beq
	\frac{d\,b_{\theta}}{d\,\theta}=-\theta\frac{1}{q^2_{\theta}}\left(\frac{d\,q_{\theta}}{d\,\theta}\right)^2+\frac{\theta}{q_{\theta}}\left(\frac{d^2\,q_{\theta}}{d^2\,\theta}\right).
	\eeq
	
	From \eqref{eq:qderiv} and \eqref{eq:qderiv2}, we can write
	\begin{multline}
	\frac{d b_{\theta}}{d \theta}=-\frac{\theta}{q_{\theta}}\left(\frac{1}{\sqrt{M}}\phi_{\infty}(x^*-\sqrt{M}\mu_{\gamma(\theta)})\left(\frac{-d \mu_{\gamma(\theta)}}{d \theta}\right)\right)^2\\
	\times\left(\frac{1}{q_{\theta}}\phi_{\infty}(x^*-\sqrt{M}\mu_{\gamma(\theta)})+\frac{1}{\sqrt{M}}(x^*-\sqrt{M}\mu_{\gamma(\theta)})\right)\\
	+\frac{\theta}{q_{\theta}}\frac{1}{M\sqrt{M}}\phi_{\infty}(x^*-\sqrt{M}\mu_{\gamma(\theta)})\left(\frac{-d^2 \mu_{\gamma(\theta)}}{d^2 \theta}\right).
	\end{multline}
	From Lemma~\ref{thm:gauss} and the conditions $\frac{d\mu_{\theta}}{d\theta}<0$,
	$\frac{d^2\mu_{\theta}}{d^2\theta}>0$, we have $\frac{d b_{\theta}}{d \theta}<0$. Therefore, there exists a unique critical point denoted by $\theta^*$.

	 Now the condition that $\frac{d r_{\theta}}{d \theta}>0$ is equivalent to $b_{\theta}>0$. Since $b_{\theta}$ is monotonically decreasing with $\theta$ we must have $\theta<\theta^*$ to satisfy the condition $
	\frac{d r_{\theta}}{d \theta}>0$. Similarly, we can show that $
	\frac{d r_{\theta}}{d \theta}<0$ for $\theta>\theta^*$. 
	
	Finally, by using the result that the unique critical point $\theta^*$ satisfies $b_{\theta}=0$, we obtain
	\eqref{eq:critical_mean}.
	This completes the proof. \qed
\end{proof}

We now study the case of $L_{\theta}=ae^{-\theta}$.
\begin{theorem}
	For $L_{\theta}=ae^{-\theta}$, if $\mu_{\theta}$ is strictly convex, then there exists a unique critical point for $r_{\theta}$ denoted by $\theta^*$. Furthermore, $r_{\theta}$ increases with $\theta$ up to $\theta*$ and then it decreases with $\theta$.
	The critical point $\theta^*$ satisfies
	\begin{align}
		\label{eq:critical_mean2}
		q_{\theta^*}&=-\frac{\phi_{\infty}(x^*-\sqrt{M}\mu_{\gamma(\theta^*)})\frac{d \mu_{\gamma(\theta^*)}}{d \theta}}{\sqrt{M}\log(\frac{q_{\theta^*}}{q_{0}})}.
	\end{align}
\end{theorem}
\begin{proof}
	
	By using $L_{\theta}=ae^{-\theta}$ and \eqref{eq:rderiv2}, we obtain that a critical point satisfies
	\beq
	-ae^{-\theta}\log\left(\frac{q_{\theta}}{q_{0}}\right)+\frac{ae^{-\theta}}{ q_{\theta}}\left(\frac{d q_{\theta}}{d \theta}\right)=0.
	\eeq
	After simplifications we obtain
	\beq
	-\log\left(\frac{q_{\theta}}{q_{0}}\right)+\frac{1}{q_{\theta}}\left(\frac{d q_{\theta}}{d \theta}\right)=0.
	\eeq
	Let $b_{\theta}$ be defined as
	\beq
	b_{\theta}=-\log\left(\frac{q_{\theta}}{q_{0}}\right)+\frac{1}{q_{\theta}}\left(\frac{d q_{\theta}}{d \theta}\right).
	\eeq
	Then every critical point satisfies $b_{\theta}=0$. We will show that $\frac{d\,b_{\theta}}{d\,\theta}<0$ which implies that there exists a unique solution to $b_{\theta}=0$. It can be verified that
	\beq
	\frac{d\,b_{\theta}}{d\,\theta}=-\frac{1}{q_{\theta}}\frac{d\,q_{\theta}}{d\,\theta}-\frac{1}{q^2_{\theta}}\left(\frac{d\,q_{\theta}}{d\,\theta}\right)^2+\frac{1}{q_{\theta}}\left(\frac{d^2\,q_{\theta}}{d^2\,\theta}\right).
	\eeq
	
	From \eqref{eq:qderiv} and \eqref{eq:qderiv2}, we can write
	\begin{multline}
		\frac{d b_{\theta}}{d \theta}=-\frac{1}{q_{\theta}}\frac{1}{\sqrt{M}}\phi_{\infty}(x^*-\sqrt{M}\mu_{\gamma(\theta)})\left(\frac{-d \mu_{\gamma(\theta)}}{d \theta}\right)
		\\
		-\frac{1}{q_{\theta}}\left(\frac{1}{\sqrt{M}}\phi_{\infty}(x^*-\sqrt{M}\mu_{\gamma(\theta)})\left(\frac{-d \mu_{\gamma(\theta)}}{d \theta}\right)\right)^2\\
		\times\left(\frac{1}{q_{\theta}}\phi_{\infty}(x^*-\sqrt{M}\mu_{\gamma(\theta)})+\frac{1}{\sqrt{M}}(x^*-\sqrt{M}\mu_{\gamma(\theta)})\right)\\
		+\frac{1}{q_{\theta}}\frac{1}{M\sqrt{M}}\phi_{\infty}(x^*-\sqrt{M}\mu_{\gamma(\theta)})\left(\frac{-d^2 \mu_{\gamma(\theta)}}{d^2 \theta}\right).
	\end{multline}
	From Lemma~\ref{thm:gauss} and the conditions $\frac{d\mu_{\theta}}{d\theta}<0$,
	$\frac{d^2\mu_{\theta}}{d^2\theta}>0$, we have $\frac{d b_{\theta}}{d \theta}<0$. Therefore, there exists a unique critical point denoted by $\theta^*$.

	Now the condition that $\frac{d r_{\theta}}{d \theta}>0$ is equivalent to $b_{\theta}>0$. Since $b_{\theta}$ is monotonically decreasing with $\theta$ we must have $\theta<\theta^*$ to satisfy the condition $
	\frac{d r_{\theta}}{d \theta}>0$. Similarly, we can show that $
	\frac{d r_{\theta}}{d \theta}<0$ for $\theta>\theta^*$. 
	
	Finally, by using the result that the unique critical point $\theta^*$ satisfies $b_{\theta}=0$, we obtain
	\eqref{eq:critical_mean}.
	This completes the proof. \qed
\end{proof}

\section{Numerical Results}
\label{sec:numerics}
In this section, we provide some numerical results to support our analysis. We choose the following parameters for our simulations: $A=1.5$, $K=15$, $\alpha=0.1$, $\phi_{\infty}$ is the pdf of $\mc{N}(0,1)$, and $\phi_{\theta}$ is the pdf of $\mc{N}(\mu_{\theta},1)$. We choose $\theta$ to be in the range $[0.1,  1.5]$ so that $L_{\theta}\leq K$ and $\theta\leq A$.

\begin{figure}
	\centering
	\includegraphics[height=0.7\columnwidth]{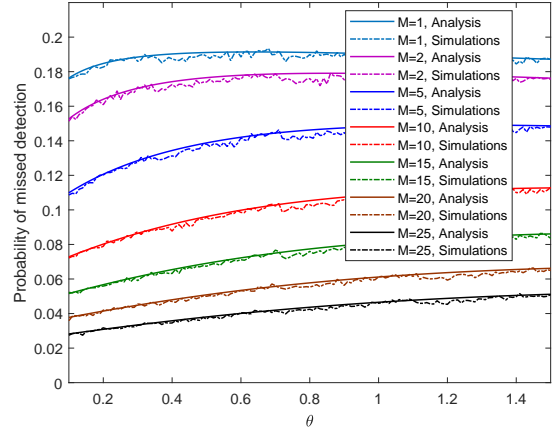}
	\caption{$Q_{\theta}(K,M,A)$ \text{ v/s } $\theta$, $L_{\theta}=L_{\theta}^{(1)}$, $\mu_{\theta}=\mu_{\theta}^{(1)}$}
	\label{pmd_mc1}
\end{figure}

\begin{figure}
	\centering
	\includegraphics[height=0.7\columnwidth]{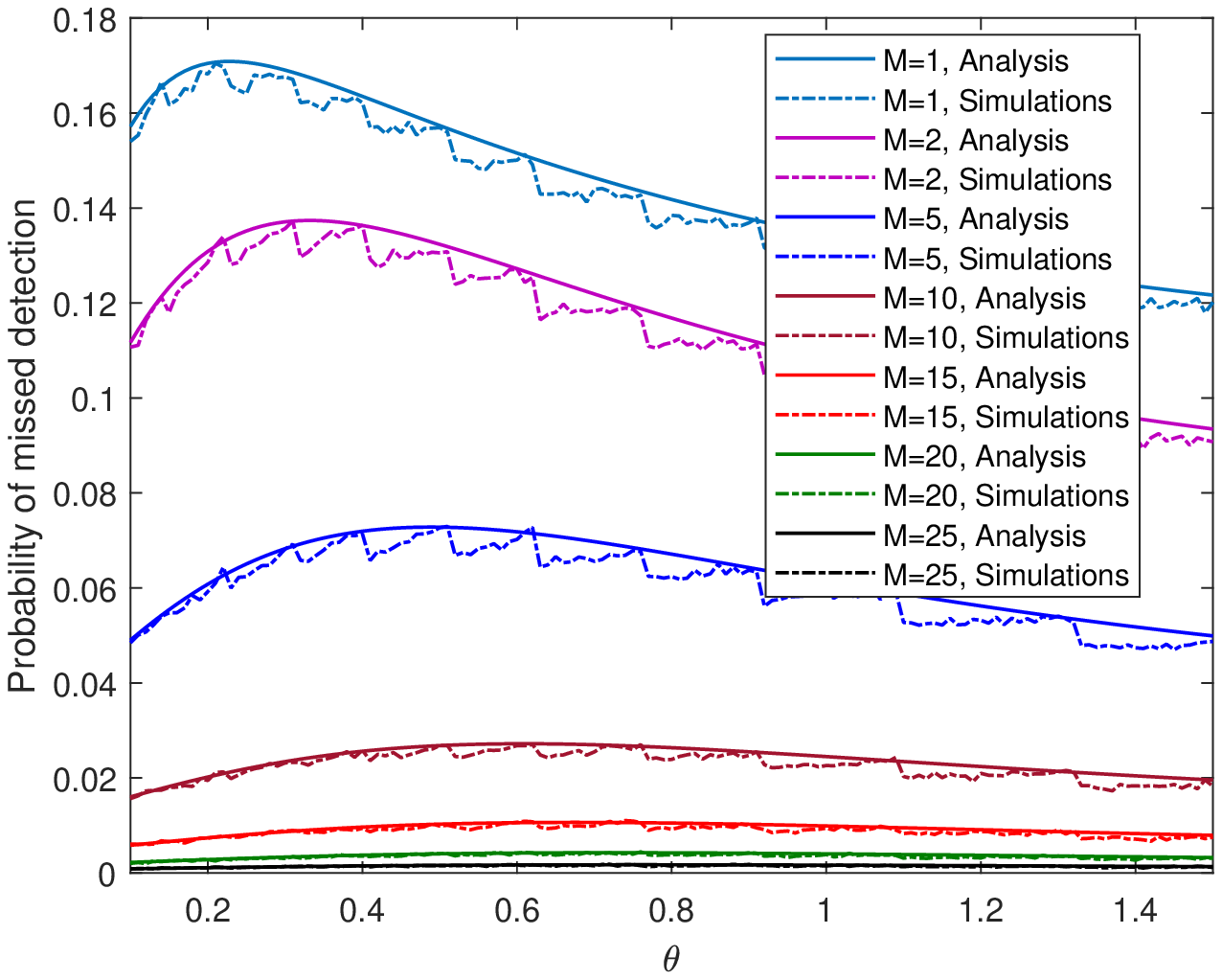}
	\caption{$Q_{\theta}(K,M,A)$ \text{ v/s } $\theta$, $L_{\theta}=L_{\theta}^{(1)}$, $\mu_{\theta}=\mu_{\theta}^{(2)}$}
	\label{pmd_mc2}
\end{figure}

\begin{figure}
	\centering
	\includegraphics[height=0.7\columnwidth]{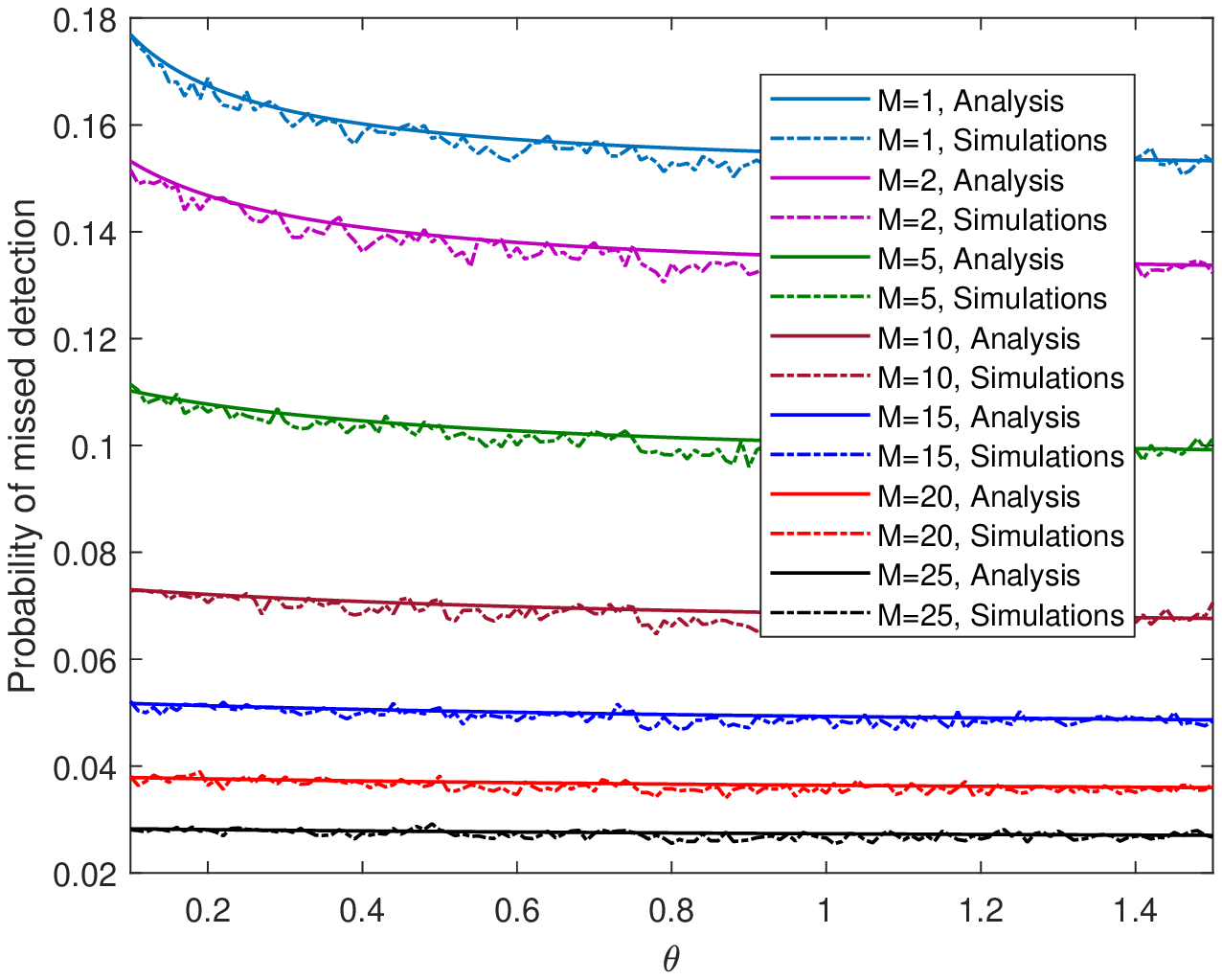}
	\caption{$Q_{\theta}(K,M,A)$ \text{ v/s } $\theta$, $L_{\theta}=L_{\theta}^{(2)}$, $\mu_{\theta}=\mu_{\theta}^{(1)}$}
	\label{pmd_mc3}
\end{figure}

\begin{figure}
	\centering
	\includegraphics[height=0.7\columnwidth]{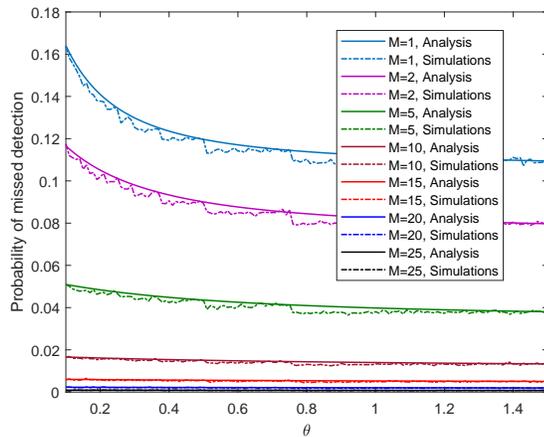}
	\caption{$Q_{\theta}(K,M,A)$ \text{ v/s } $\theta$, $L_{\theta}=L_{\theta}^{(2)}$, $\mu_{\theta}=\mu_{\theta}^{(2)}$}
	\label{pmd_mc4}
\end{figure}

We consider the following two different functions for $\mu_{\theta}$: $\mu_{\theta}^{(1)}=\frac{0.1}{1+10\theta}$ and $\mu_{\theta}^{(2)}= 0.2e^{-10\theta}$. For $L_{\theta}$, we choose two types of functions:
$L_{\theta}^{(1)}=10A e^{-\theta}$ and $L_{\theta}^{(2)}=\frac{A}{\theta}$.
In Figures~\ref{pmd_mc1}-\ref{pmd_mc4}, we plot $Q_{\theta}(K,M,A)$ as a function of $\theta$ for different values of $M$. We observe that there exists a unique $\theta$ that maximizes $Q_{\theta}(K,M,A)$. Furthermore, the PMD decreases with the number of sensors.
 From Figure~\ref{pmd_mc1}, we observe that the detector has to choose $M=25$ to make the worst-case PMD less than $\delta=0.05$ for the case $L_{\theta}=L_{\theta}^{(1)}$ and $\mu_{\theta}=\mu_{\theta}^{(1)}$. Similarly, we can find the number of sensors required to make the PMD less than 0.05 for the remaining cases.

%
%
%
%
%
%


\section{Conclusions}
\label{sec:conclusions}
We considered a scenario where an adversary manipulates sensors' observations to reduce the probability of its detection within a certain number of time-slots after the changepoint. For the Gaussian mean-change problem, under the assumption that the detector uses the Shewhart stopping test, we showed the existence of a unique optimal parameter for the adversary to maximize the PMD. We then find sufficient number of sensors to be used such that the worst-case PMD is less than an acceptable level.  In future work, it is of interest to extend the analysis to the setting where the detector can use different types of sensors and has budget constraints on resources that it can spend on sensors.
\bibliographystyle{IEEEbib}
\bibliography{refs}
\end{document}